\documentclass[english,aps,superscriptaddress,manuscript]{revtex4-1}
\usepackage[T1]{fontenc}
\usepackage[latin9]{inputenc}
\usepackage{float}
\usepackage{amsmath}
\usepackage{amssymb}
\usepackage{graphicx}

\makeatletter

\providecommand{\tabularnewline}{\\}

 \@ifundefined{textcolor}{}
 {%
   \definecolor{BLACK}{gray}{0}
   \definecolor{WHITE}{gray}{1}
   \definecolor{RED}{rgb}{1,0,0}
   \definecolor{GREEN}{rgb}{0,1,0}
   \definecolor{BLUE}{rgb}{0,0,1}
   \definecolor{CYAN}{cmyk}{1,0,0,0}
   \definecolor{MAGENTA}{cmyk}{0,1,0,0}
   \definecolor{YELLOW}{cmyk}{0,0,1,0}
 }

\usepackage{babel}

\makeatother

\usepackage{babel}
\begin{document}

\title{New IceCube data and color octet neutrino interpretation of the PeV
energy events}

\author{A. N. Akay}

\email{aakay@etu.edu.tr}

\selectlanguage{english}%

\address{TOBB University of Economics and Technology, 06560, Ankara, Turkey.}

\author{O. Cakir}

\email{ocakir@science.ankara.edu.tr}

\selectlanguage{english}%

\address{Ankara University, Faculty of Sciences, Department of Physics, 06100,
Tandogan, Ankara, Turkey.}

\author{Y. O. Gunaydin}

\email{yogunaydin@ksu.edu.tr}

\selectlanguage{english}%

\address{Kahramanmaras Sutcu Imam University, Faculty of Sciences and Letter,
Department of Physics, Kahramanmaras, Turkey}

\author{U. Kaya}

\email{ukaya@etu.edu.tr}

\selectlanguage{english}%

\address{TOBB University of Economics and Technology, 06560, Ankara, Turkey.}

\author{M. Sahin}

\email{mehmet.sahin@usak.edu.tr}

\selectlanguage{english}%

\address{Usak University, Faculty of Sciences and Literature, Department of Physics,
64200, Usak, Turkey.}

\author{S. Sultansoy}

\email{ssultansoy@etu.edu.tr}

\selectlanguage{english}%

\address{TOBB University of Economics and Technology, 06560, Ankara, Turkey.}

\address{National Academy of Sciences, Institute of Physics, Baku, Azerbaijan.}
\begin{abstract}
IceCube collaboration has published two papers on ultrahigh energy
neutrinos observation, recently. They have used the data collected
in two years in their first publication, which reveals observation
of two PeV energy neutrino events. The second publication of the collaboration
including more data has also confirmed main features of the former
paper. In literature, various interpretations of the IceCube data
have been proposed. In this study, it is shown that PeV energy neutrino
events observed by the IceCube collaboration can be interpreted as
resonance production of color octet neutrinos with masses in $500-800$
GeV range. 
\end{abstract}
\maketitle

\section*{1. Introduction}

Observation of two $PeV$ energy neutrino events by the IceCube collaboration
\cite{Aartsen2013} has led to plenty of papers on the possible interpretations
of these events (see review \cite{Anchordoqui2014} and references
therein). As mentioned in Ref. \cite{Aartsen2013}: `` The events
were discovered in a search for ultrahigh energy neutrinos using data
corresponding to $615.9$ days effective live time and probability
of observing two or more candidate events under the atmospheric background-only
hypothesis is $2.9\times10^{-3}$ ($2.8\sigma$)''.

Recent publication \cite{Aartsen2014}, which includes three years
of IceCube data, has confirmed main features of previous two years
data \cite{AartsenScience} (for discussion of IceCube data see Section
2): concerning $PeV$ region one more events is observed, and there
are no events between $0.4\: PeV$ and $1\: PeV$, as well as above
$2.5\: PeV$. These features may follow either from `` mono-energetic''
neutrino source in $PeV$ region or from $TeV$ scale resonance in
neutrino-nucleon scattering (or both). The gap in $0.4\div1.0\: PeV$
region certainly means that excesses observed in $1\div2.5\: PeV$
and $0.03\div0.4\: PeV$ regions have different origins.

In Section 2 we present a brief summary of the IceCube results, emphasizing
the gap between $0.4\: PeV$ and $1\: PeV$ as well as zero result
for energies above $2.5\: PeV$. Color octet neutrino ($\nu_{8}$)
phenomenology is considered in Section 3. Section 4 is devoted to
$\nu_{8}$ interpretation of $PeV$ neutrino events. In Section 5
leptoquark interpretation of $PeV$ events is reconsidered taking
into account new IceCube data. Finally we give some concluding remarks
in Section 6.

\section*{2. New IceCube data}

As mentioned in previous section first two $PeV$ events were observed
by IceCube within $615.9$ days effective live time. Results of two-year
dataset {[}4{]}, namely, $28$ high-energy neutrino events within
$662$ days effective live time, are grouped in \cite{Anchordoqui2014}
as following (there is a misprint in the first item, $50\: TeV$ should
be replaced by $30\: TeV$): 

$\bullet\,26$ events from $50\: TeV$ to $1\: PeV$, which includes
the $\sim10$ atmospheric background events; 

$\bullet\,2$ events from $1\: PeV$ to $2\: PeV$; 

$\bullet\,$zero events above $2\: PeV$, say from $2\: PeV$ to $10\: PeV$,
with a background of zero events.

In a similar manner, results of three-year dataset \cite{Aartsen2014},
corresponding to 988 days effective live time, can be grouped as:

$\bullet\:33$ events from $30\: TeV$ to $0.4\: PeV$, which includes
the $\sim10$ atmospheric background events;

$\bullet\:$zero events from $0.4\: PeV$ to $1\: PeV$, with a background
of $0.2$ events; 

$\bullet\,3$ events from $1\: PeV$ to $2.5\: PeV$, with a background
of $0.02$ events;

$\bullet\:$zero events above $2.5\: PeV$, say from $2.5\: PeV$
to $10\: PeV$, with a background of zero events.

As mentioned in the Introduction, these features may be followed from: 

$\bullet\,$\textquotedbl{}mono-energetic\textquotedbl{} neutrino
source(s) in $PeV$ region (see \cite{Anchordoqui2014} and references
therein), i.e. blazars or gamma ray bursts (GRB's); 

$\bullet\,$or from $TeV$ scale resonance in neutrino-nucleon scattering,
namely, leptoquarks or leptogluons (first interpretation has been
considered in \cite{Barger} and second one in \cite{Akay2014}); 

$\bullet\,$or both. 

\noindent \begin{flushleft}
Let us repeat that the gap in $0.4\div1.0\: PeV$ region certainly
means that excesses observed in $1\div2.5\: PeV$ and $0.03\div0.4\: PeV$
regions have different origins.
\par\end{flushleft}

All the three $PeV$ energy events (event ID's $14$, $20$ and $35$
in \cite{Aartsen2014}) has shower type event topology. Two events
observed in first two years have deposited energies $1041_{-144}^{+132}\: PeV$
and $1141_{-133}^{+143}\: PeV$. Third event observed during third
year has deposited energy $2004_{-262}^{+236}\: PeV$.

\section*{3. Phenomenology of color octet neutrinos}

Color octet (decuplet) neutrinos and leptons as well as color sextet
($15-plet$) quarks are predicted by preonic models with colored preons
\cite{Souza,Celikel1998,Hewett1997,Sultansoy1998}. There are two
strong arguments favoring preon models: inflation of ``fundamental\textquotedblright{}
particles and free parameters in the SM (other BSM models, including
SUSY, drastically increase the number of free parameters) and mixing
of ``fundamental\textquotedblright{} quarks and leptons. The first
one, namely ``inflation\textquotedblright{}, historically results
in discovery of new level of matter two times during the last century:
periodical table of chemical elements was clarified by Rutherford
experiment, inflation of hadrons results in quark model (see Table
1 from \cite{Sahin2011}).

Recently, color octet leptons have been come to forefront again \cite{Sahin2010,Akay2011,Mandal,Netto,Sahin2014}.
It should be noted that concerning preon models color octet neutrinos
has the same status as color octet leptons, and the status of both
of them is similar to exited leptons and neutrinos, which are widely
investigated by ATLAS and CMS collaborations \cite{Aad2013ATLAS,Aad2014ATLAS,Chatrhyan2014CMS,Khachatryan2014CMS}.

\textbf{\textit{Fermion-scalar and three-fermion models (see \cite{Sultansoy1998}
and references therein):}}

Keeping a minimal scheme in mind, we make two assumptions: i) There
is no parastatistics, ii) Preons are colored objects. According to
the first assumption the SM fermions should contain odd number of
fermionic preons, which lead to fermion-scalar models or three fermion
models. The second assumption means that preons are color triplets.

\textit{Leptons}: In the framework of fermion-scalar models, leptons
would be a bound state of one fermionic preon and one scalar anti-preon

\[
l=(F\overline{S})=1\oplus8
\]

\noindent \begin{flushleft}
then each SM lepton has one colour octet partner. In a three fermion
model, the color decomposition
\par\end{flushleft}

\[
l=(FFF)=1\oplus8\oplus8\oplus10
\]

\noindent \begin{flushleft}
predicts the existence of two color octet and one color decouplet
partners.
\par\end{flushleft}

\textit{Quarks:} In fermion-scalar models, anti-quarks are consist
of one fermionic and one scalar preons which means that each SM anti-quark
has one colored sextet partner
\[
\overline{q}=(FS)=\overline{3}\oplus6
\]

\noindent \begin{flushleft}
According to the three fermion models
\par\end{flushleft}

\[
q=(F\overline{F}F)=3\oplus\overline{3}\oplus\overline{6}\oplus15
\]

\noindent \begin{flushleft}
therefore, for each SM quark one anti-triplet, one anti-sextet and
one 15-plet partners are predicted.
\par\end{flushleft}

In this paper we consider color octet neutrinos in the framework of
fermion-scalar models. The interaction lagrangian for color octet
neutrinos is given by

\begin{equation}
L=\frac{1}{2\Lambda}\underset{l}{\sum}\{\bar{v}_{l8}^{a}g_{s}G_{\mu\nu}^{a}\sigma^{\mu\nu}(\eta_{L}\nu_{lL}+\eta_{R}\nu_{lR})+h.c.\}.
\end{equation}

\noindent \begin{flushleft}
Here, $\Lambda$ is compositeness scale, $G_{\mu\nu}^{a}$ is field
strength tensor for gluon, index $a=1,2,...,8$ denotes the color,
$g_{s}$ is the gauge coupling, $\eta_{L}$ and $\eta_{R}$ are the
chirality factors, $\nu_{lL}$ and $\nu_{lR}$ denote left and right
spinor components of neutrino ($l=e,$$\mu,\tau$), $\sigma^{\mu\nu}$
is the antisymmetric tensor. According to neutrino chirality conservation,
$\eta_{L}\eta_{R}=0$. We set $\eta_{R}=0$ in this analysis.
\par\end{flushleft}

According to PDG \cite{PDG} current exclusion limit for color octet
neutrino is $M_{\nu_{8}}>110$$GeV$ assuming $\nu_{8}\rightarrow\nu+g$
decay. This value is obtained from Tevatron data, a rough estimation
shows that $M_{\nu_{8}}$ below $400\, GeV$could be excluded by current
LHC data.

\section*{4. Color octet neutrino as the source of IceCube PeV events}

Feynman diagram for resonant $\nu_{8}$ production in $\nu N$ scattering
is given in Figure 1.

\begin{figure}[H]
\begin{centering}
\includegraphics[scale=0.5]{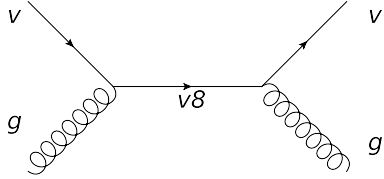} 
\par\end{centering}

\caption{Resonant $\nu_{8}$ production in the ultra high energy neutrino nucleon
scattering.}
\end{figure}

In order to perform numerical calculations we implement color octet
neutrino interaction lagrangian, given in Eq. 1, into the CalcHEP
software \cite{CALCHEP}. In Fig. 2 we present $\nu_{8}$ production
cross-section as a function of incoming cosmic neutrino energy for
different $\nu_{8}$ mass values. For numerical calculations we set
$\Lambda=m_{\nu_{8}}$ together with $CTEQ6L$ \cite{CTEQ6L} parton
distributions. It should be noted that resonance cross-section is
proportional to $(\eta_{L}/\Lambda)^{2}$ .

\begin{figure}[H]
\begin{centering}
\includegraphics[scale=0.75]{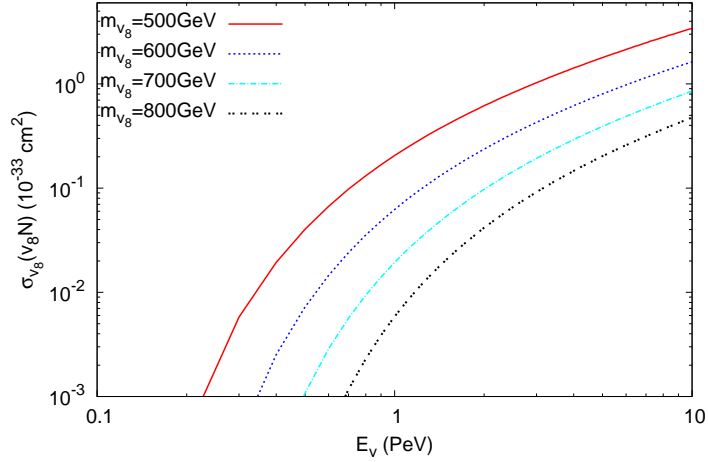} 
\par\end{centering}

\caption{$\nu_{8}$ production cross section in $\nu N$ scattering.}
\end{figure}

In order to obtain $\nu_{8}$ production rate distribution the cross-section
given in Fig. 2 should be convoluted with ultra high energy neutrino
flux. According to \cite{Aartsen2014} the best fit power law for
extraterrestrial neutrino flux is

\begin{equation}
E^{2}\phi(E)=1.5\times10^{-8}(E/100\: TeV)^{-0.3}\: GeV\: cm^{-2}\: s^{-1}\: sr^{-1}
\end{equation}

\noindent \begin{flushleft}
for each neutrino species. For the calculation of production rate
distribution we use
\par\end{flushleft}

\begin{equation}
dN_{\nu_{8}}/dE_{\nu}=nt\Omega\sigma_{\nu_{8}}\phi(E)
\end{equation}

\noindent \begin{flushleft}
where $t=988$ days is the time exposure, $n=6\times10^{38}$ is the
effective number of target nucleons in IceCube and $\Omega=4\pi$.
\par\end{flushleft}

In $\nu_{8}$ interpretation, there are three possible sources of
$PeV$ events namely $\nu_{e_{8}}$, $\nu_{\mu_{8}}$ and $\nu_{\tau_{8}}$.
Real picture depends on $\nu_{8}$ mass hierarchy. Below we assume
that masses of color octet neutrinos are close to each other. Production
rate distribution $dN_{\nu_{8}}/dE_{\nu}$ is given in Fig. 3.

\begin{figure}[H]
\begin{centering}
\includegraphics[scale=0.75]{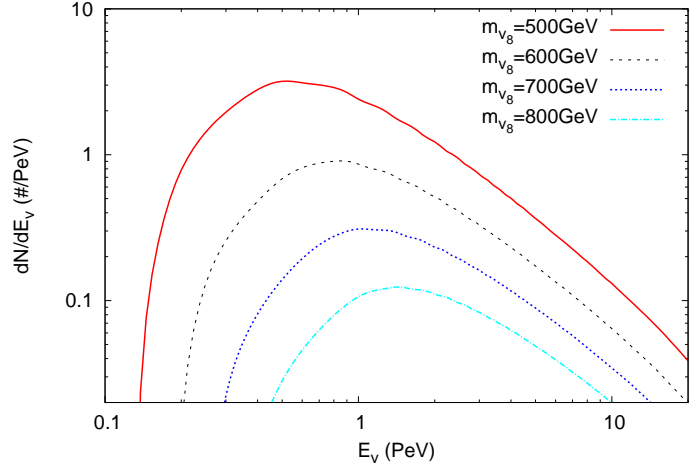} 
\par\end{centering}

\caption{Production rate distribution $dN_{\nu_{8}}/dE_{\nu}$ from the $\nu_{8}$
cross-section convoluted with extraterrestrial neutrino flux. }
\end{figure}

Up to this stage, procedure is similar to leptoquark case that considered
in \cite{Barger} and rough estimations for $\nu_{8}$ case performed
in \cite{Akay2014}. Concerning $\nu_{8}$ interpretation, in fact
essential part of energy is carried by neutrino, therefore it is invisible,
and IceCube shower is formed by gluon. For this reason gluon rate
distribution, presented in Fig. 4, is more appropriate for IceCube
data analysis.

\begin{figure}[H]
\begin{centering}
\includegraphics[scale=0.75]{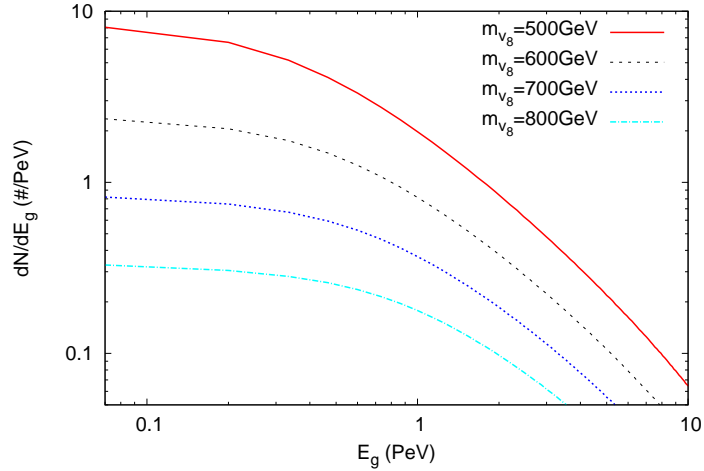} 
\par\end{centering}

\caption{Event rate distribution $dN_{\nu_{8}}/dE_{g}$ from the gluon energy
distribution convoluted with extraterrestrial neutrino flux. }
\end{figure}

Let us turn to interpretation of IceCube $PeV$ energy events. As
mentioned in Section 2, there are $3$ events in $1-2.5\: PeV$ region,
whereas there are no events in $0.4-1\: PeV$ region and above $2.5\: PeV$.
In our analysis, we use $3$ observed events as a basic point. Then
referring to Fig. 2 from \cite{Aartsen2014}, we use $2.5\:(3.5)$
events as upper limit for $0.4-1\:(2.5-10)\: PeV$ regions. In Table
1 we present expected number of events in different energy regions
normalized to $3$ events observed in $1-2.5\: PeV$ region. Last
column contains values of $\eta_{L}$ which provides exactly $3$
events in $1-2.5\: PeV$ region.

\begin{table}[H]
\caption{Number of events for different masses of color octet neutrinos}

\centering{}%
\begin{tabular}{|c|c|c|c|c|c|}
\hline 
$m_{\nu_{8},}GeV$  & $0-0.4\: PeV$  & $0.4-1.0\: PeV$  & $1.0-2.5\: PeV$  & $2.5-10\: PeV$  & $\eta_{L}$\tabularnewline
\hline 
\hline 
$500$  & $4.8$  & $2.3$  & $3.0$  & $2.4$  & $0.4$\tabularnewline
\hline 
$600$  & $3.4$  & $2.2$  & $3.0$  & $2.7$  & $0.6$\tabularnewline
\hline 
$700$  & $2.4$  & $2.0$  & $3.0$  & $3.0$  & $0.9$\tabularnewline
\hline 
$800$  & $1.9$  & $1.9$  & $3.0$  & $3.3$  & $1.2$\tabularnewline
\hline 
\end{tabular}
\end{table}

It is seen that color octet neutrinos with mass values between $500$
and $800\: GeV$ provide correct interpretation of IceCube data, whereas
masses above $800\: GeV$ lead to excess of events above $2.5\: PeV$.
Mass value corresponding to $\eta_{L}=1$ is approximately $740\: GeV$.

\section*{5. Comments on leptoquark interpretations of $PeV$ events}

Leptoquark interpretation of two $PeV$ events observed within first
2 years IceCube data \cite{Aartsen2013} was proposed in \cite{Barger}.
As the result, authors mention that scalar leptoquark of charge $-1/3$,
which couples to the first generation quarks and the third generation
leptons, with a mass around $600\: GeV$ and coupling $f_{L}=1$ provides
correct description of the data. In this section we reconsider their
results using new (3 years) IceCube data. Using Fig. 3 from \cite{Barger}
we estimate number of events for different mass values and energy
regions. Similar to Table 1 we use $3$ observed PeV energy events
as a basic point. Results are presented in Table 2.

\begin{table}[H]
\caption{Number of events for different masses of leptoquarks}

\centering{}%
\begin{tabular}{|c|c|c|c|c|c|}
\hline 
$m_{\nu_{LQ},}GeV$  & $0-0.4\: PeV$  & $0.4-1.0\: PeV$  & $1.0-2.5\: PeV$  & $2.5-10\: PeV$  & $f_{L}$\tabularnewline
\hline 
\hline 
$500$  & $2.5$  & $6.4$  & $3.0$  & $1.2$  & $1.0$\tabularnewline
\hline 
$600$  & $0.5$  & $3.9$  & $3.0$  & $1.1$  & $1.3$\tabularnewline
\hline 
$700$  & $0$  & $2.1$  & $3.0$  & $1.0$  & $1.7$\tabularnewline
\hline 
$800$  & $0$  & $1.1$  & $3.0$  & $0.9$  & $2.2$\tabularnewline
\hline 
\end{tabular}
\end{table}

It is seen that leptoquark interpretation is in conflict with new
IceCube data if $m_{\nu_{LQ}}$ is below $650\:\, GeV$, because it
leads to excess of events in $0.4-1\: PeV$ energy region. Leptoquark
with mass above $650\: GeV$ is in agreement with data, however coupling
constant $f_{L}$ should be larger than $1.5$.

\section*{6. Conclusion}

Observation of $PeV$ energy neutrino events in the IceCube experiment
may lead to serious consequences both for astrophysics and particle
physics. In the light of our calculations, color octet neutrinos with
masses in $500-800\: GeV$ range give rational interpretation of the
IceCube results on $PeV$ energy neutrinos. In addition, scalar leptoquark
with mass $650-700\: GeV$ also provide reasonable mechanisms for
IceCube data. Correctness of both interpretations may be checked in
near future using forthcoming IceCube and LHC data.


\begin{thebibliography}{10}
\bibitem[1]{Aartsen2013} M. G. Aartsen et al. {[}IceCube Collaboration{]},
Phys. Rev. Lett. 111, 021103 (2013); arXiv:1304.5356 {[}astroph.HE{]}.

\bibitem[2]{Anchordoqui2014} L. A. Anchordoqui et al., Journal of
High Energy Astrophysics 1-2 (2014) 1-30; arXiv:1312.6587v3 {[}astro-ph.HE{]}.

\bibitem[3]{Aartsen2014} M. G. Aartsen et al. {[}IceCube Collaboration{]},
Phys. Rev. Lett. 113 (2014) 101101; arXiv:1405.5303 {[}astroph.HE{]}.

\bibitem[4]{AartsenScience} M. G. Aartsen et al. {[}IceCube Collaboration{]},
Science 342, 1242856 (2013); arXiv:1311.5238v2 {[}astro-ph.HE{]}.

\bibitem[5]{Barger} V. Barger and W. -Y. Keung, Phys. Lett. B 727,
190(2013); arXiv:1305.6907 {[}hep-ph{]}.

\bibitem[6]{Akay2014} A.N. Akay, U. Kaya and S. Sultansoy; arXiv:1402.1681 {[}hep-ph{]}.

\bibitem[7]{Souza} I.A D'Souza, C. S. Kalman, PREONS: Models of Leptons,
Quarks and Gauge Bosons as Composite Object,World Scientific Publishing
Co., (1992).

\bibitem[8]{Celikel1998} A. Celikel and M. Kantar, Turk. J. Phys.
22, 401 (1998).

\bibitem[9]{Hewett1997} J. L. Hewett and T. G. Rizzo, Phys. Rev.
D 56, 5709 (1997); arXiv:hep-ph/9703337.

\bibitem[10]{Sultansoy1998} A. Celikel, M. Kantar, and S. Sultansoy,
Phys. Lett. B 443, 359 (1998).

\bibitem[11]{Sahin2011} M. Sahin, S. Sultansoy, S. Turkoz, Phys.
Rev. D 83, 054022 (2011); arXiv:1009.5405 {[}hep-ph{]}.

\bibitem[12]{Sahin2010} M. Sahin, S. Sultansoy, and S. Turkoz, Phys.
Lett. B 689, 172 (2010); arXiv:1001.4505 {[}hep-ph{]}.

\bibitem[13]{Akay2011} A. N. Akay, H. Karadeniz, M. Sahin, and S.
Sultansoy, Europhys. Lett. 95, 31001 (2011); arXiv:1012.0189 {[}hep-ph{]}. 

\bibitem[14]{Mandal} T. Mandal and S. Mitra, Phys. Rev. D 87, 095008
(2013); arXiv:1211.6394 {[}hep-ph{]}.

\bibitem[15]{Netto} D. G. Netto, D. L. Val, K. Mawatari, I. Wigmore,
and T. Plehn, Phys. Rev. D 87, 094023 (2013); arXiv:1303.0845 {[}hep-ph{]}.

\bibitem[16]{Sahin2014} M. Sahin, Acta Phys. Pol. B 45 (2014) 1811-1831;
arXiv:1302.5747 {[}hep-ph{]}.

\bibitem[17]{Aad2013ATLAS} G. Aad et al. (ATLAS Collaboration), New
J. Phys. 15 (2013) 093011; arXiv:1308.1364 {[}hep-ex{]}.

\bibitem[18]{Aad2014ATLAS} G. Aad et al. (ATLAS Collaboration), Phys.
Lett. B 728 (2014) 562-578; arXiv:1309.3230 {[}hep-ex{]}.

\bibitem[19]{Chatrhyan2014CMS} S. Chatrhyan et al. (CMS Colloboration),
Phys. Lett. B 720 (2013) 309; arXiv:1210.2422 {[}hep-ex{]}.

\bibitem[20]{Khachatryan2014CMS} V. Khachatryan et al. (CMS Colloboration),
arXiv:1406.5171 {[}hep-ex{]}.

\bibitem[21]{PDG} J. Beringer et al. (Particle Data Group), Phys.
Rev. D 86, 010001 (2012).

\bibitem[22]{CALCHEP} A. Belyaev, N.D. Christensen, A. Pukhov, Comput.
Phys. Commun. 184, 1729 (2013); arXiv:1207.6082 {[}hep-ph{]}.

\bibitem[23]{CTEQ6L} J. Pumplin et al., JHEP 0207, 012 (2002); arXiv:hep-ph/0201195.\end{thebibliography}
\end{document}